\documentclass[pra,twocolumn,superscriptaddress]{revtex4}
\usepackage{color}
\usepackage{amsfonts}
\usepackage{amsmath}
\usepackage{amssymb}
\usepackage{graphicx}

\begin{document}

\title{Theory of record thermopower near a finite temperature magnetic phase
transition: IrMn}
\author{Peter W\"{o}lfle }
\affiliation{Institute for Theory of Condensed Matter, Karlsruhe Institute of Technology,
76128 Karlsruhe, Germany}
\affiliation{Institute for QuantumMaterials and Technologies, Karlsruhe Institute of
Technology, 76021 Karlsruhe, Germany }
\author{Timothy Ziman}
\affiliation{Institute Laue-Langevin,  71 Avenue des Martyrs, 38042 Grenoble, France }

\begin{abstract}
The effect of scattering of conduction electrons by dynamical spin
fluctuations on the thermopower in metals near a thermal phase transition
into an antiferromagnetic phase is considered. We are interested in a
transition at room temperature, as has been studied in a heterostructure
involving layers of IrMn. We show that the electrical resistivity exhibits a
narrow but low peak at the transition, which may be  ddifficult to detect on top
of the main contributions induced by phonons and impurities. By contrast, the
thermopower is found to exhibit a prominent peak both as a function of
temperature $T$ for fixed layer thickness $t_{AFM}$ and as a function of $%
t_{AFM}$ for fixed $T.$ We conjecture that the transition temperature $T_{c}$
is a function of both $t_{AFM}$ and the Fermi energy $\epsilon _{F}$. Both
dependencies give rise to a sharp peak of the thermopower as a function
of $T$ or $t_{AFM}$ near the transition. The estimated magnitude of the peak
for the case of three-dimensional longitudinal spin fluctuations is in good
agreement with experiment.
\end{abstract}

\maketitle

\section{Introduction}

The recently observed ftacular temperature dependence of the thermopower
in a magnetic heterostructure involving the antiferromagnetic metal IrMn 
\cite{Ziman20} suggests that the spin fluctuations near the
antiferromagnetic transition may be responsible for the observed peak at the
ordering temperature. This finding is all the more interesting in that the
magnetic transition temperature may be tuned by the thickness of the IrMn
layers to be at room temperature, which makes the effect highly promising
for applications. The effect of elastic scattering of conduction electrons
by the local spins of magnetic metals near the phase transition into a
magnetically ordered phase has been studied first by DeGennes and Friedel 
\cite{DeGennes58}, who assumed the spin configuration to be temperature
dependent and static as given by an Ornstein-Zernike form. Somewhat later,
Fisher and Langer \cite{Fisher68} revisited the problem in the light of the
theory of classical critical phenomena. These authors pointed out that the
equal time spin correlation function entering the scattering cross section
also appears in the internal magnetic energy and is therefore related
to the specific heat. They also mentioned that inelastic scattering may be
important. The effect of inelastic scattering on the thermopower was
considered by Entin-Wohlman, Deutscher, and Orbach \cite{Entin76} in a model
calculation leaving the dynamics of the local spin system as an input
quantity to be determined from case to case. A detailed model calculation of
the electrical resistivity of antiferromagnetic metals, in the framework of
the Self-Consistent Renormalization Theory of spin fluctuations in itinerant
magnets \cite{Hasegawa74,Moriya85} has been worked out by Ueda \cite{Ueda77}%
. In all these previous studies the anomalies near the transition found for
the transport properties appeared to be relatively weak and cannot account
for the prominent peak found in the thermopower as a function of temperature
or as a function of layer thickness of an IrMn heterostructure \cite{Ziman20}.

In this paper we estimate the contribution of scattering of the charge
carriers by dynamical spin fluctuations by assuming a phenomenological form
of the spin excitation spectrum dictated by symmetry and conservation laws.
We observe that the presence of gapless fermionic excitations in a metal
changes the spin excitation spectrum of the local spins in a decisive way,
leading to a strongly temperature dependent, at $T_{c}$ seemingly divergent
contribution to the resistivity. The divergent behavior arises for not too
high transition temperature, $T_{c}\ll \epsilon _{F}$, where $\epsilon _{F}$
is the Fermi energy (here and in the following we use energy units of
Kelvin). The divergence is cut off close to $T_{c}$ at a temperature $T_{x}$%
, marking the transition into a quantum critical regime.\ The effect may be
characterized as a continuation of quantum critical scattering in a narrow
region of the phase diagram along the phase boundary. There are two effects
introduced by the itinerant electrons into the spin dynamics of \
antiferromagnetic metals as contrasted to insulators. First, the transition
temperature is shifted by an amount proportional to the static wave vector
dependent conduction electron spin susceptibility at the ordering wave
vector $\mathbf{Q}$, $\chi _{s}(\mathbf{Q,}0)$. The latter depends on the
Fermi energy and thus gives rise to a dominant contribution to the
thermopower. Secondly, the dynamics of the spin fluctuations at low energy
is dominated by the Landau damping mechanism leading to inelastic scattering
processes strongly enhanced near the transition.

\section{Model and Method}

\subsection{Hamiltonian}

We assume a system of interacting localized spins, coupled to conduction
electrons as expressed by the Hamiltonian

\begin{equation}
H=H_{c}+H_{S}+H_{ex}
\end{equation}
where $H_{c}$ represents a single conduction band

\begin{equation}
H_{c}=\sum_{\mathbf{k},\alpha }\epsilon _{\mathbf{k}}c_{\mathbf{k}\alpha
}^{\dag }c_{\mathbf{k}\alpha },
\end{equation}%
We assume the conduction electron system to be three-dimensional.

The dynamics of the localized spin system is defined by

\begin{equation}
H_{S}=\sum_{\mathbf{q}}\sum_{\alpha =x,y,z}I_{\alpha }S_{\mathbf{q}}^{\alpha
}S_{-\mathbf{q}}^{\alpha },
\end{equation}%
where we allow for anisotropic interaction $I_{z}\gg I_{x,y}$ caused by the
strong spin-orbit interaction at the Ir ions.

The coupling of conduction electrons to localized spins is described by

\begin{equation}
H_{c-S}=J\sum_{\mathbf{q}}\mathbf{s}_{\mathbf{q}}\mathbf{S}_{-\mathbf{q}}.
\end{equation}%
Here $S_{\mathbf{q}}^{\alpha }$ are the Fourier components of the localized
spin operators, and the conduction electron spin operator is defined by $%
\mathbf{s}_{\mathbf{q}}=\sum_{\mathbf{k},\alpha ,\beta }\mathbf{\tau }%
_{\alpha \beta }c_{\mathbf{k+q}\alpha }^{\dag }c_{\mathbf{k}\beta }$ , with $%
\mathbf{\tau }_{\alpha \beta }$ the vector of Pauli matrices (the coupling
constants $I_{\alpha },J$ are given in units of [energy/density]).

\subsection{Spin fluctuations in the paramagnetic phase}

The transport properties of the conduction electron system of MnIr in the
temperature range around room temperature are governed by electron-phonon
interaction and the exchange interaction mediated by $H_{c-S}$ as we argue
below. We assume the system to be anisotropic in spin space, with preferred
direction along the $z-$axis. We will consider both three-dimensional and
two-dimensional spin fluctuations, where the $3d$ model appears to describe
the experiment \cite{Ziman20} better, as we shall see.

In the absence of coupling of local spins and conduction electron spins, for 
$J=0$, the longitudinal susceptibility of localized spins $\chi _{loc}^{zz}$
is assumed to be well approximated by the static Ornstein-Zernike form

\begin{equation}
\chi _{loc}^{zz(0)}(\mathbf{q,}\omega )\approx \frac{n_{S}}{I_{z}}\frac{1}{%
(T-T_{c,I})/T_{c,I}+(\mathbf{q-Q})^{2}\xi _{0}^{2}}
\end{equation}%
where $\mathbf{Q}$ is the ordering wave vector and $\xi _{0}$ is a
microscopic spin interaction length of the order of a lattice spacing and $%
n_{S}$ is the density of localized spins. The transition to the
antiferromagnetic phase is signaled by the divergence of $\chi _{loc}^{zz(0)}
$ at wave vector $\mathbf{q=Q}$ and at temperature $T_{c,I}=O(I_{z})$. In
the quasi two-dimensional slab geometry of IrMn within the nanostructure
studied in \cite{Ziman20} the transition temperature is found to depend on
the layer thickness $t_{AFM}$. One source of such a dependence is the
quantization of the momentum component $(\mathbf{q-Q})_{z}$ normal to the
layer surface, $(\mathbf{q-Q})_{z}=\frac{\pi }{t_{AFM}}(2n+1)$, $n=0,\pm
1,...$. The minimal value of $(\mathbf{q-Q})_{z}$ at $n=0$ gives rise to a
suppression of the transition temperature 
\begin{equation}
T_{c,I}^{\ast }=T_{c,I}[1-(\frac{\pi \xi _{0}}{t_{AFM}})^{2}]
\label{Tc-tAFM}
\end{equation}%

\begin{figure}[tbp]
\includegraphics[width=1.15\columnwidth]{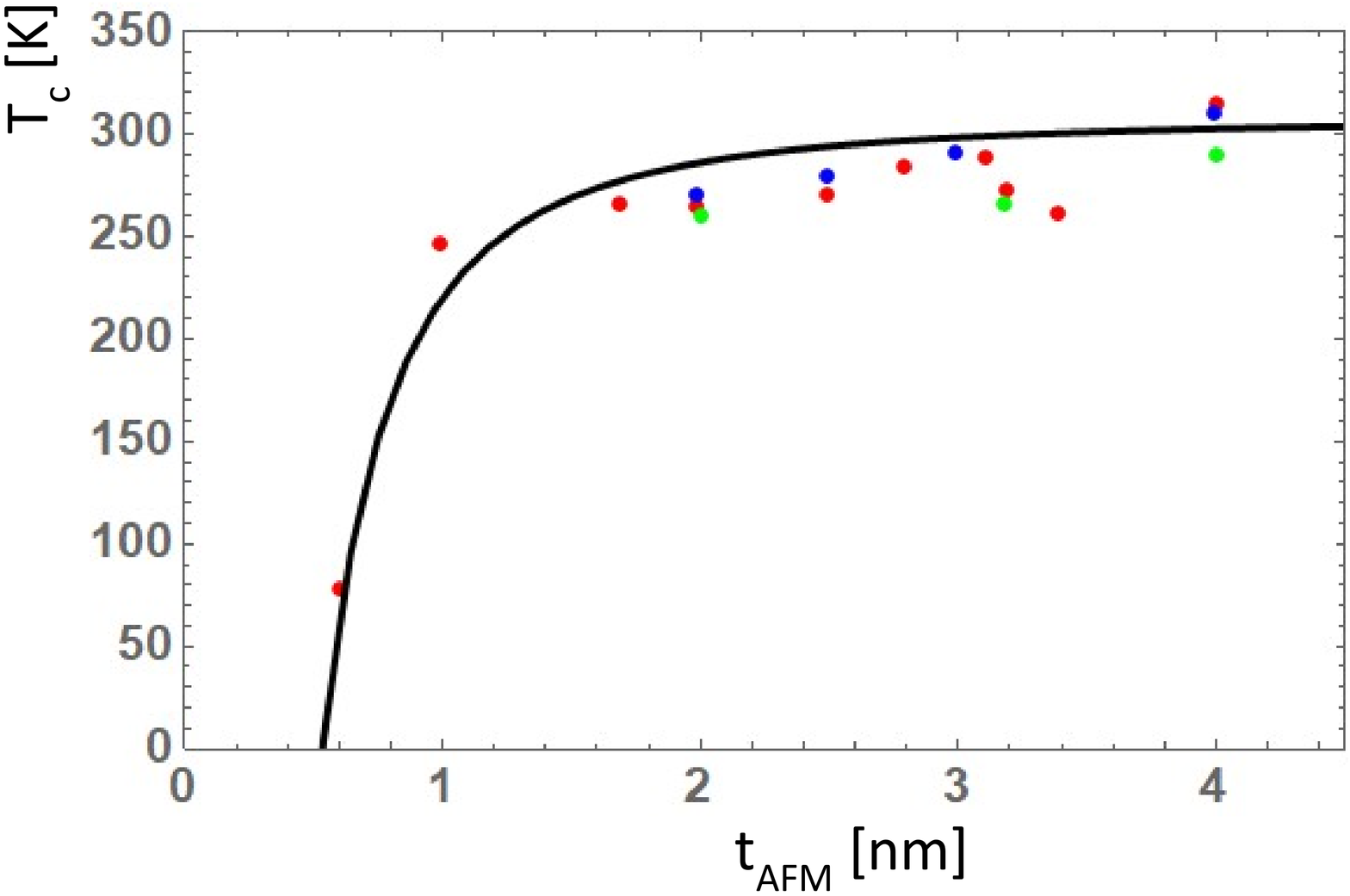}
\caption{Transtion temperature into the antiferromagnetic phase versus 
layer thickness $t_{AFM}$. Theory according to Eq.~({\protect\ref{Tc-tAFM})} 
(solid line). Data points from \protect\cite{Ziman20}: Seebeck measurements (red dots); 
susceptibility data (blue dots); XMLD data (green dots).}
\label{fig:Tc-tAFM}
\end{figure}

$\ $In Fig.\ \ref{fig:Tc-tAFM} we show that the $T_{c}$ data of Fig. 4 of \cite%
{Ziman20} may be fitted reasonably well by  Eq.~({\ref{Tc-tAFM}) using a spin
interaction length }$\xi _{0}=0.17$nm and $T_{c,I}^{\ast }=308$K. Here the
reference transition temperature $T_{c,I}^{\ast }$ is not the bulk value,
which is much higher, but a reduced value appropriate for the composite
layer structure. For example, the effective spin exchange energy (the
quantity $I_{z}$ in our notation) has been found to be strongly varying with
the Fermi energy \cite{Umetsu04},  possibly accounting for a reduction of
the bulk $T_{c}$ value by a factor of $\sim 3$.  The above estimate differs
somewhat from the theoretical results for $T_{c}$ versus $t_{AFM}$ given in 
\cite{Frangou16} and the model calculation for the suppression of $T_{c}$ based
on the reduction of the interaction volume of a spin near the surface \cite%
{Zhang01}.

The conduction electron spin susceptibility in the model of noninteracting
fermions is given at low energy by

\begin{equation}
\chi _{s}(\mathbf{q,}\omega )\approx \chi _{s}(\mathbf{q,}0)[1+i\pi \frac{%
\omega }{v_{F}q}]
\end{equation}%
where $v_{F}$ is the Fermi velocity (we employ an isotropic band structure
for simplicity) and we assume isotropy in spin space.\ In the presence of
coupling the susceptibility of localized spins is renormalized approximately
as

\begin{equation}
\chi _{loc}^{zz}(\mathbf{q,}\omega )\approx \frac{1}{[\chi _{loc}^{zz(0)}(%
\mathbf{q,}\omega )]^{-1}+J^{2}\chi _{s}(\mathbf{q,}\omega )}
\end{equation}%
In the vicinity of the transition and for small $\omega $ the susceptibility 
$\chi _{loc}^{zz}$ is sharply peaked at $\mathbf{q\approx Q}$ so that $\chi
_{s}(\mathbf{q,}\omega )$ may be replaced by $\chi _{s}(\mathbf{Q,}\omega )$%
, and $\chi _{loc}^{zz}$ takes the form

\begin{equation}
\chi _{loc}^{zz}(\mathbf{q,}\omega )\approx \frac{n_{S}}{I_{z}}\frac{1}{\tau
+(\mathbf{q-Q})^{2}\xi _{0}^{2}+\frac{i\omega }{\gamma }}
\end{equation}%
Here we defined the reduced temperature $\tau =(T/T_{c}-1)$ and a
renormalized transition temperature $T_{c}=T_{c,I}^{\ast }-\Delta T_{c}$
where $\Delta T_{c}=T_{c,I}^{\ast }n_{S}J^{2}\chi _{s}(\mathbf{Q,}%
0)/I_{z}\approx T_{c,I}^{\ast }(n_{S}J^{2}/I_{z})N_{F}$, with $%
N_{F}=mk_{F}/\pi ^{2}$ the conduction electron density of states at the
Fermi level ($m$ is the mass and $k_{F}$ is the Fermi wave number of the
conduction electrons). The transition temperature $T_{c}$ is seen to depend
on the Fermi energy $\epsilon _{F}=k_{F}^{2}/2m$ , a fact of considerable
importance for the thermopower as we show later. The reference energy $%
\gamma $ entering the Landau damping term in the denominator of $\chi
_{loc}^{zz}(\mathbf{q,}\omega )$ is defined as $\gamma
=[(n_{S}J^{2}/I_{z})\chi _{s}(\mathbf{Q,}0)]^{-1}v_{F}Q/\pi \approx
(I_{z}/\pi n_{S}J^{2}N_{F})v_{F}Q\approx T_{c}(N_{F}J)^{-2}$, assuming $%
Q\approx k_{F}$ and $n_{S}\approx n\approx N_{F}\epsilon _{F}.$

\subsection{Spin fluctuations in the antiferromagnetic phase}

In the ordered phase the translation invariance is broken as indicated by
the ordering vector $\mathbf{Q}$. The rotation invariance in spin space is
also broken by the appearance of the ordered moment vector $\mathbf{M(R}%
_{j}) $ . Spin fluctuations along $\mathbf{M}$ (longitudinal) and orthogonal
to $\mathbf{M}$ (transverse) behave in a fundamentally different way. The
magnetic order also affects the quasiparticle excitations in major ways. The
Brillouin zone gets shrunk into the magnetic Brillouin zone, leading to
multiple bands induced by back-folding. To keep things simple we will assume
two-sublattice antiferromagnetic order ($\mathbf{Q=(}\pi ,\pi ,\pi )$) for
which case one finds two bands, dubbed "valence" and "conduction" band,
separated by a gap $\Delta $. The structure of the quasiparticle and
collective excitations has been calculated for the Hubbard model within the
Random Phase Approximation \cite{Moriya12,Schrieffer89,Chubukov92}.

\subsubsection{Longitudinal spin fluctuations}

The longitudinal spin susceptibility is finite in the ordered phase but
diverges upon approach to the thermal transition. This is expressed by the
form

\begin{equation}
\chi _{loc}^{zz}(\mathbf{q},\omega )\approx \frac{n_{S}}{I_{z}}\frac{1}{%
r_{l}+(\mathbf{q-Q})^{2}\xi _{0}^{2}+i(\omega /\gamma )},
\end{equation}%
where $r_{l}=c_{l}M_{s}^{2}=c_{l}M_{0}^{2}|\tau |$, assuming mean field
behavior of the ordered moments $M_{s}\propto M_{0}|\tau |^{1/2}$, where $%
M_{0}$ is the saturated moment. Here $c_{l}$ is a constant of $O(1)$.

\subsubsection{Transverse spin fluctuations}

\ If spin rotation invariance in the plane orthogonal to the ordered moments 
$\mathbf{M(R}_{j})$ is still preserved, which we suspect not to be the case
for MnIr, gapless spin excitations (spin waves) exist. In this case the
transverse spin susceptibility involving spin excitations orthogonal to $%
\mathbf{M(R}_{j})$ is divergent in the whole ordered phase, not just at the
critical point. More generally, the transverse spin susceptibility may be
expressed as  
\begin{equation}
\chi ^{-+}(\mathbf{q},\omega )\approx \frac{n_{S}}{I_{z}}\frac{1}{\Delta
_{s}/\gamma +(\mathbf{q-Q})^{2}+i(\omega /\gamma )}
\end{equation}%
where $\Delta _{s}$ is the spin wave gap.\ The coupling of the transverse
spin fluctuations to the quasiparticles depends on the respective
quasiparticle bands. For definiteness we assume that the Fermi energy lies
in the lower (valence) band. Then the scattering of quasiparticles from the valence
band into the valence band is suppressed, the corresponding vertex $\Lambda
_{vv}^{tr}(\mathbf{k,k+q+Q)}\propto \mathbf{q\cdot \nabla }\epsilon _{%
\mathbf{k}}$ such that the product $(\Lambda _{vv}^{tr})^{2}\chi ^{-+}$
appearing in the expression for the self energy is no longer singular and
hence does not give rise to critical behavior \cite%
{Schrieffer89,Chubukov92,Adler65}.\ By contrast, the vertex function for
scattering from the valence band into the conduction band is of $O(1)$.
However, the latter excitation \ requires a minimum energy $\Delta $ and is
thus possible only in the vicinity of the transition when $T\gtrsim \Delta $%
. In the following we will assume that the spin wave gap $\Delta _{s}>T_{c}$
that spin waves may not be thermally excited.

\section{Electrical resistivity \ }

We now present a model calculation of the contribution of scattering by
antiferromagnetic fluctuations to the electrical resistivity $\varrho $.
Since the typical momentum transfer in such scattering processes is large,
of order of the ordering wave vector $\mathbf{Q}$, and therefore, assuming a
half-filled conduction band, of order $k_{F}$,\ the momentum relaxation rate 
$1/\tau _{tr}$ is approximately related to the imaginary part of the
electron self energy $\Sigma (\mathbf{k,}\omega )$ by $1/\tau _{tr}\approx 2%
\mathit{Im} \Sigma $. The resistivity is then given in terms of the self energy
as%
\begin{equation}
\rho \approx \frac{m}{e^{2}n}2\mathit{Im}\Sigma (k_{F}\mathbf{,}\omega \approx
T)
\end{equation}

\subsection{Three-dimensional spin fluctuations}

In a clean metal at low temperature scattering by antiferromagnetic
fluctuations affects only a small part of the Fermi surface: the "hot spots"
connected by the ordering wave vector $\pm \mathbf{Q}$. At higher
temperatures phonon scattering, or at all temperatures impurity scattering,
helps to remove the constraints imposed by momentum conservation so that the
critical behavior at the hot spots is distributed all over the Fermi
surface. The self-energy may be calculated to one-loop order, taking into
account phonon and impurity scattering, which leads to a prefactor $A=O(1)$.

In the case of electrons and spin fluctuations that are both three-dimensional, we get

\begin{eqnarray}
\mathit{Im}\Sigma (\mathbf{k,}\omega ) &\sim &AJ^{2}\int_{-\infty }^{\infty }%
\frac{d\nu }{2\pi }\int \frac{d^{3}q}{8\pi ^{3}}\mathit{Im} G(\mathbf{k+q+Q}%
,\omega +\nu )  \notag \\
&&\times \mathit{Im}\chi (\mathbf{q+Q,}\nu )[f(\nu +\omega )+b(\nu )]
\end{eqnarray}%
where $f(\omega ),b(\nu )$ are the Fermi and Bose functions respectively, and $J$ is the
coupling constant of electrons and spin fluctuations. The conduction
electron spectral function is approximated by

\begin{equation}
\mathit{Im} G(\mathbf{k},\omega)=\pi\delta(\omega-\epsilon_{\mathbf{k}})
\end{equation}
We will employ an isotropic model of the conduction band with energy $%
\epsilon_{k}=k^{2}/2m-\epsilon_{F}$, for which the angular integral may be
done, assuming $q\ll Q$, with the result

\begin{equation}
\frac{1}{2}\int_{-1}^{1}d\cos \theta \mathit{Im} G(\mathbf{k+q+Q},\omega +\nu
)\approx \frac{\pi }{2}\frac{m}{k_{F}Q}\approx \frac{1}{\epsilon _{F}}
\end{equation}%
At low frequency $\omega <T$ we may drop $\omega $ in the argument of the
Fermi function, whence $\mathit{Im} \Sigma (\mathbf{k,}\omega )$ is
approximately independent of frequency

\begin{eqnarray}
\mathit{Im}\Sigma (\mathbf{k,}\omega  &\lesssim &T)\approx A\frac{n_{S}J^{2}}{%
I_{z}\epsilon _{F}}\int_{-\infty }^{\infty }\frac{d\nu }{2\pi \sinh (\nu /T)}
\\
&&\times \int_{0}^{\infty }\frac{dqq^{2}}{2\pi ^{2}}\frac{\nu /\gamma }{%
(\tau +(q/k_{F})^{2})^{2}+(\nu /\gamma )^{2}}  \notag
\end{eqnarray}

The divergence of the momentum integral at $q\rightarrow 0$ is cut off in
two ways: (i) in the limit $\tau \rightarrow 0$ the Landau damping term
provides the cutoff just as is the case at the quantum critical point (QCP).
The ensuing quantum critical behavior is therefore found to extend in the
phase diagram from the QCP along the phase boundary in a narrow strip of
width $\tau _{x}T_{c}$, where $\tau _{x}\approx T_{c}/\gamma $; (ii) in the
limit $\nu \rightarrow 0$, or more generally $T\rightarrow 0$ (considering
that the frequency integral is confined to $|\nu |\lesssim T$) the cutoff is
provided by the term $\tau $ . This is the quantum disordered regime, in
which $\mathit{Im} \Sigma $ is found to depend critically on $\tau $. The
latter behavior may be seen in analogy to the approach to the QCP at $T=0$
from the quantum disordered side. 

Provided the transition temperature is not too high, $T_{c}\ll \gamma $ ,
there exists a wide regime of reduced temperatures $\tau _{x}\ll \tau \ll 1$
with $\tau _{x}\approx T_{c}/\gamma $ \ for which $\mathit{Im}\Sigma $ is
approximately given by

\begin{align}
\mathit{Im}\Sigma (\mathbf{k,}\omega & \lesssim T)\approx \frac{3}{2\pi }A%
\frac{n_{S}nJ^{2}T}{I_{z}\epsilon _{F}\gamma }\int_{0}^{T}d\nu \int_{\sqrt{%
\tau }}\frac{dq/k_{F}}{(q/k_{F})^{2}}  \notag \\
& \approx \frac{3}{2\pi }A\frac{n_{S}nJ^{2}}{I_{z}}\frac{T^{2}}{\epsilon
_{F}\gamma }\frac{1}{\sqrt{\tau }}
\end{align}%
where $n=k_{F}^{3}/3\pi ^{2}$ is the density of conduction electrons. The
resistivity follows as

\begin{equation}
\rho _{sfl}^{3d}=\rho _{0}\frac{1}{\sqrt{\tau }},\text{ \ \ }\tau _{x}\ll
\tau \text{ \ }  \label{rho3}
\end{equation}%
where 
\begin{equation}
\rho _{0}\approx \frac{9}{4}R_{Q}k_{F}^{-1}A\frac{n_{S}J}{I_{z}}\frac{nJ}{%
\epsilon _{F}}\frac{T_{c}^{2}}{\epsilon _{F}\gamma },\text{ \ }
\end{equation}
and $R_{Q}=\frac{h}{e^{2}}\approx 25.81$k$\Omega $ is the quantum
resistance. In the ordered phase, in the temperature regime $\tau _{x,l}\ll
|\tau |\ll 1$ , where $\tau _{x,l}\approx T_{c}/\gamma _{l}$, with $\gamma
_{l}=\gamma c_{l}M_{0}^{2}$, the resistivity scale $\rho _{0}$ in Eq.~({\ref%
{rho3})} \ is replaced by $\rho _{0,l}=\rho _{0}/(M_{0}\sqrt{c_{l}})$. In
the temperature regime close to the transition, defined by $-\tau
_{x,l}<\tau <\tau _{x}$ , the resistivity is given by 
\begin{equation}
\rho _{sfl}^{3d}=\rho _{0}\frac{1}{\sqrt{\tau _{x}}},\text{ \ \ \ }-\tau
_{x,l}<\tau <\tau _{x}
\end{equation}

We now estimate the $\rho _{sfl}$ of IrMn layers as studied in \cite{Ziman20}%
. The reference resistance setting the scale is given by $%
R_{Q}k_{F}^{-1}\approx 258\mu \Omega $cm taking $k_{F}\approx 10^{8}$cm$%
^{-1} $. The remaining factors at the transition temperature are $\gamma
\approx 10^{4}$K, $T_{c}/\gamma \approx 0.03$ , $I_{z}\approx T_{c}\approx
300$K, $nJ\approx N_{F}J\epsilon _{F}\approx 1.5\times 10^{3}$K, assuming $%
\epsilon _{F}\approx 10^{4}$K,\ and $A\approx 1$ , resulting in $\rho
_{0}\approx 0.4\mu \Omega $cm. This is very small in comparison to the
observed resistivity (Fig.2 of the Supplementary information of Tu et al. 
\cite{Ziman20}\ where $\rho \approx 170\mu \Omega $cm at $T_{c}$ ). Indeed,
the resistivity data do not show any trace of a peak at $T_{c}$. Here we
anticipate that the crossover scale $\tau _{x}\approx 0.06$ (see below),
leading to the estimate of the maximum height of the spin fluctuation
induced contribution as $\rho _{sfl}(T_{c})=\rho _{0}\frac{1}{\sqrt{\tau _{x}%
}}\approx 1.5\mu \Omega $cm. The model calculations for bulk MnIr \cite%
{Umetsu04} show a hump in the electrical resistivity below $T_{c}$ .

\subsection{Two-dimensional spin fluctuations}

In this case the self-energy at low frequency is found as

\begin{align}
\mathit{Im} \Sigma (\mathbf{k,}\omega & \lesssim T)\approx \frac{A}{2\pi ^{2}}%
\frac{n_{S}J^{2}}{I_{z}\gamma }\int_{-\infty }^{\infty }\frac{d\nu \nu
/\gamma }{\sinh (\nu /T)}  \notag \\
& \times \int \frac{dqq/t_{AFM}}{(\tau +(q/k_{F})^{2})^{2}+(\nu /\gamma )^{2}%
} \\
& \approx \frac{3}{2}A\frac{J^{2}nn_{S}}{I_{z}}\frac{T^{2}}{\gamma \epsilon
_{F}}\frac{1}{k_{F}t_{AFM}}\frac{1}{\tau },
\end{align}%
using $n=k_{F}^{3}/3\pi ^{2}$. We note that magnetic order is not destroyed
by transverse spin fluctuations because those are gapped out. The
resistivity shows a more strongly divergent behavior for $\tau \rightarrow 0$%
, up to the crossover temperature $\tau _{x}$

\begin{equation}
\rho _{sfl}^{2d}=\rho _{0}\frac{\pi }{k_{F}t_{AFM}}\frac{1}{\tau }\left\{ 
\begin{array}{c}
1,\text{ \ \ \ \ \ \ \ \ \ }\tau \gg \tau _{x} \\ 
\frac{1}{c_{l}M_{0}^{2}},\text{\ \ }-\tau \gg \tau _{x,l}%
\end{array}%
\right. 
\end{equation}%
The maximum of $\rho _{sfl}^{2d}$ is reached for $\tau <\tau _{x}$ and may
be estimated as $\rho _{sfl}^{2d}\approx \rho _{0}\frac{\pi }{k_{F}t_{AFM}}%
\frac{1}{\tau _{x}}\approx 0.5\mu \Omega $cm, assuming $t_{AFM}=30$\AA ,
which is again\ much less than the total resistivity.

The system showing a record thermopower studied in \cite{Ziman20} consists
of layers of IrMn of thickness varying between $0.6-4$nm and is therefore
quasi two-dimensional. The momentum component $q_{z}$ of the spin
fluctuations normal to the sample plane is quantized, $q_{z,n}=\frac{\pi }{%
t_{AFM}}(2n+1)$, $n=0,1,2...,$ where $t_{AFM}$ is the sample thickness. For
sufficiently small thickness $t_{AFM}$ only the lowest transverse mode is
occupied. For this to be valid we should have $(\frac{2\pi }{t_{AFM}}\xi
_{0})^{2}\gg \tau _{x}$, where $\xi _{0}\approx k_{F}^{-1}$ is the
microscopic spin interaction length. For the IrMn layers of thickness $%
\approx 3$nm the latter condition is not satisfied, taking $\xi _{0}\approx
0.17$nm. We therefore conclude that the three-dimensional model is more
appropriate for describing the sample with $t_{AFM}=2.8$nm, for which a
detailed comparison with the thermopower data is possible (see below).\ 

\section{Thermopower}
The thermopower is defined by

\begin{equation}
S=-\frac{\pi ^{2}k_{B}^{2}T}{e}\frac{\rho ^{\prime }(\epsilon _{F})}{\rho
(\epsilon _{F})}
\end{equation}%
where $\rho ^{\prime }(E_{F})=d\rho /d\epsilon _{F}\approx $ $d\rho
_{sfl}/d\epsilon _{F}$, assuming that of all contributions to the
resistivity, by scattering off phonons, off impurities, off magnetic ions,
the largest contribution is coming from critical spin fluctuations. The
essential dependence of $\rho _{sfl}$ on $\epsilon _{F}$ is through the
shift of the transition temperature $T_{c}$ \ induced by changing $\epsilon
_{F}$ . Using $d\tau /d\epsilon _{F}=-(T/T_{c}^{2})dT_{c}/d\epsilon _{F}$.
one then finds above the transition in case of three- \ or two-dimensional
spin fluctuations 

\begin{equation}
S=\frac{\pi ^{2}k_{B}^{2}}{e}\frac{\rho _{0}}{\rho }\frac{dT_{c}}{d\epsilon
_{F}}(\frac{T}{T_{c}})^{2}\Theta (\tau -T_{c}/\gamma )\left\{ 
\begin{array}{c}
\tau ^{-3/2}\text{, \ \ \ \ }d=3 \\ 
\frac{\pi }{k_{F}t_{AFM}}\tau ^{-2}\text{,\ \ }d=2%
\end{array}%
\right. ,
\end{equation}%
a power law divergence $S\propto (T-T_{c})^{-\alpha }$, which is cut off at $%
\tau \approx T_{c}/\epsilon _{F}$. The dependence of \ $T_{c}$ on $\epsilon
_{F}$\ has been discussed above, $dT_{c}/d\epsilon _{F}\approx
(n_{S}J/I_{z})(JN_{F})(T_{c}/\epsilon _{F})\approx (N_{F}J)^{2}\approx 0.02$%
. The peak height follows, using these estimates and $k_{B}/e\approx
8.6\times 10^{-3}$V/K , as $S(\tau _{x})\approx 240\mu $V/K, taking $\tau
_{x}=0.06$, \ for both three- or two-dimensional fluctuations, which is of
the order of magnitude observed in experiment, at least for the compounds
containing only IrMn magnetic layers. The compounds consisting of additional
adjacent CoFeB magnetic layers show even stronger thermopower, possibly
because the coupling constants $J,I$ are effectively changed by the magnetic
environment (increased $J$, decreased $I_{z}$). 
For temperatures below the transition the
prefactor $\rho _{0}$ and the cutoff scale $\tau _{x}$\ are replaced by $%
\rho _{0,l}$ and $\tau _{x,l}$.

\begin{figure}[tbp]
\includegraphics[width=1.1\columnwidth]{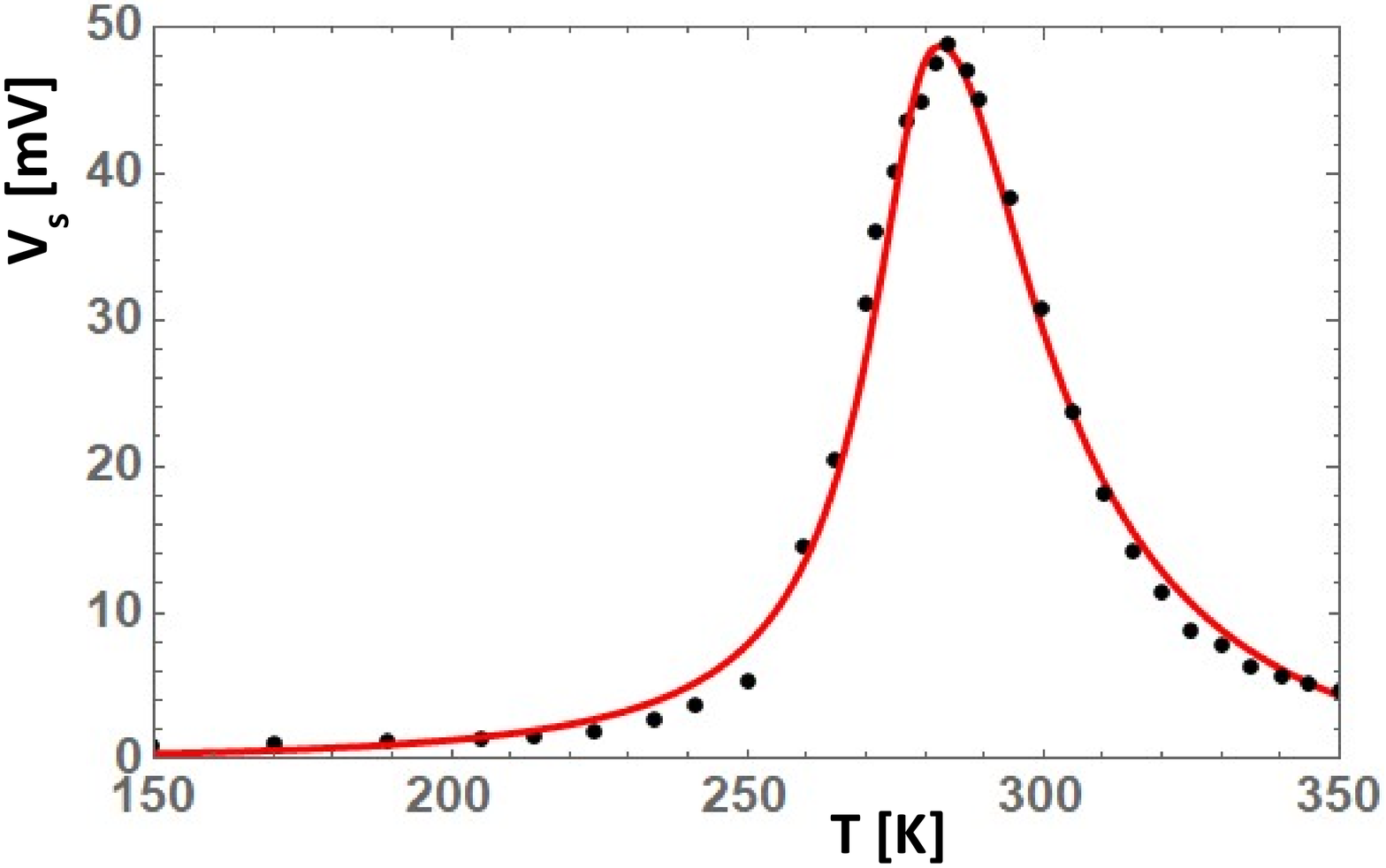}
\caption{Seebeck voltage of a IrMn nanostructure of thickness $t_{AFM}=2.8$%
nm versus temperature. Theory according to Eq.~({\protect\ref{VS3})} (solid
line). Data points from \protect\cite{Ziman20}. The transition temperature
to the antiferromagnetic phase is fixed at $T_{c}=281$K.}
\label{fig:VS-T}
\end{figure}

\ In Fig.\ \ref{fig:VS-T} we show a
comparison of our theory with the Seebeck voltage data of a sample with IrMn
layer thickness of $t_{AFM}=2.8$nm. The transition temperature is chosen to
be $T_{c}=281$K, slightly less than the $285$K estimated in \cite{Ziman20}.
We use an interpolation expression connecting the quantum critical regime at 
$|\tau |<\tau _{x}$ and the quantum disordered regime at $|\tau |>\tau _{x}$
at $T>T_{c}$: $|\tau |\Theta (\tau _{x}-|\tau |)+\tau _{x}\Theta (\tau
_{x}-|\tau |)\approx \sqrt{\tau ^{2}+\tau _{x}^{2}}$ (here $\Theta (x)$ is
the unit step function). A similar expression is employed for $T<T_{c}$. We
find that the best fit is obtained assuming three-dimensional spin
fluctuations ($d=3$). The Seebeck voltage $V_{S}$ at $T\gtrless T_{c}$ is
then described by 
\begin{equation}
V_{S}^{>,<}=c_{1}^{>,<}(\frac{T}{T_{c}})^{2}[(\frac{T}{T_{c}}-1)^{2}+(\frac{%
T_{x}^{>,<}}{T_{c}}-1)^{2}]^{-0.75}+c_{2}^{>,<}  \label{VS3}
\end{equation}%
The parameters used for fitting Eq.~({\ref{VS3}) to the data of \ Fig.2 of }%
\cite{Ziman20} are $T_{x}^{>}=299$K, $T_{x}^{<}=269.5$K, $c_{1}^{>}=0.89$mV, 
$c_{1}^{<}=0.4$mV, $c_{2}^{>}=-6.5$mV, and $c_{2}^{<}=0$mV. The constants $%
c_{2}^{>,<}$ account for contributions derived from other scattering
processes (phonons, impurities, magnetic ions) or from $d\rho _{0}/d\epsilon
_{F}$.

\section{Conclusion}

Usually the spin fluctuations in a spin system near a continuous thermal
phase transition  into a magnetically ordered state are classical. In a
metal, however, the dynamics may have quantum character even at finite
temperature on account of coupling of the localized spins to the conduction
electron spins, giving rise to Landau damping. Provided the transition
temperature is not too high, $T_{c}\ll \gamma $, where $\gamma $ is the
characteristic energy scale of the Landau damping, quantum fluctuations may
give rise to quantum critical behavior. As we have shown here, the
electrical resistivity acquires a quantum critical contribution following a
power law divergence in the reduced temperature $|\tau |^{-\alpha }$ up to a
crossover scale $\tau _{x}\ll 1$. While the corresponding peak structure in
the resistivity may be small in comparison to the dominant terms due to
scattering by phonons, magnetic moments or impurities, it gives rise to a
prominent peak in the thermopower. This is because the critical temperature $%
T_{c}$ is found to depend sufficiently strongly on the Fermi energy. We
compare our theoretical results with recent experimental observations \cite%
{Ziman20} on a heterostructure containing the antiferromagnetic metal IrMn
in the form of layers of several nanometers thickness, and find excellent
agreement. The strong dependence of the critical temperature on the layer
thickness found in  \cite{Ziman20} may also be explained by our theory. 
\par While we have focussed on the experiments on IrMn, the results suggest that large values of thermopower may result in other metallic films from the coupling to magnetic fluctuations. The current experimental results  argue for fluctuations that are still  in the three-dimensional regime, both from the form of Figure\ \ref{fig:Tc-tAFM}, where the reduction by finite size of $ T_{c,I}^{\ast }$ from the limit 
$ T_{c,I}$ is modest compared to the cutoff  $\tau _{x}\approx T_{c}/\gamma $ for the films of interest, and from the fit to the thermopower data in Fig. \ref{fig:VS-T}.  Note that we have argued that the much greater reduction in critical temperature from  bulk IrMn is due to a change in the microscopic parameters  ({\it e.g.} $I_z$) for the layers rather than a finite-size effect, in the language of critical phenomena. One might expect even stronger enhancement  at room temperature by effective two-dimensionality, provided  films of some material are thin enough that the relative reduction due to finite size, {\it{i.e.}} as in Equation \ref{Tc-tAFM}, is greater than the cut-off $\tau_{x}$.  For a fixed critical temperature smaller values of $\tau_{x}$ would require a larger  Landau damping parameter $\gamma$, which seems unlikely.
The effect of quantum fluctuations at a finite temperature antiferromagnetic 
transition has also been seen in specific heat data for two heavy fermion metals, 
CeCu$_{6-x}$Au$_{x}$ \cite{HvL94} and YbRh$_{2}$Si$_{2}$ \cite{Krellner09}
as will be shown in upcoming work \cite{WS21}.

\section{Acknowledgments}

We are grateful to Sadamichi Maekawa who suggested the importance of  magnetic fluctuations to thermopower.
PW acknowledges support by a Distinguished Senior Fellowship of Karlsruhe
Institute of Technology. TZ thanks the Reimei program of the ASRC, JAEA, Tokai, Japan for support.

\end{document}